\documentclass[aip,graphicx]{revtex4-1}

\usepackage{ifpdf}
\ifpdf
   \usepackage[pdftex]{graphicx}
   \newcommand{\pdf}[1]{#1} \newcommand{\eps}[1]{}
\else
   \usepackage{graphicx}
   \newcommand{\eps}[1]{#1} \newcommand{\pdf}[1]{}
\fi

\usepackage{color,amsmath,amsfonts}
\usepackage{float}

\newcommand{\topic}[1]{}

\newcommand{\T}{^{\mathsf{T}}}
\newcommand{\rmd}{\mathrm{d}}
\newcommand{\rme}{\mathrm{e}}
\newcommand{\rmx}{\mathrm{x}}
\newcommand{\dt}{{\Delta t}}
\newcommand{\bfx}{\mathbf{x}}
\newcommand{\bfy}{\mathbf{y}}
\newcommand{\bfp}{\mathbf{p}}
\newcommand{\bfz}{\mathbf{z}}
\newcommand{\bfw}{\mathbf{w}}
\newcommand{\calJ}{\mathcal{J}\!}

\newcommand{\calP}{\mathcal{P}}
\newcommand{\fa}{f^\mathrm{A}}
\newcommand{\fb}{f^\mathrm{B}}
\newcommand{\Phia}{\Phi^\mathrm{A}}
\newcommand{\Phib}{\Phi^\mathrm{B}}
\newcommand{\ps}{\pi}  
\newcommand{\ci}{K}  
\newcommand{\zee}{z}
\newcommand{\substep}{substep}

\floatstyle{ruled}
\newfloat{algorithm}{H}{loa}
\floatname{algorithm}{Algorithm}
\begin{document}


\title{Compressible Generalized Hybrid Monte Carlo} 

\author{Youhan Fang}
\email[]{yfang@purdue.edu}
\homepage[]{https://www.cs.purdue.edu/people/graduate-students/yfang/}
\affiliation{Purdue University}

\author{J.M.~Sanz-Serna}
\email[]{sanzsern@mac.uva.es}
\homepage[]{http://www.sanzserna.org}
\affiliation{Universidad de Valladolid}

\author{Robert D. Skeel}
\email[]{rskeel@purdue.edu}
\homepage[]{http://bionum.cs.purdue.edu}
\affiliation{Purdue University}

\date{\today}

\begin{abstract}
One of the most demanding calculations is to generate random samples
from a specified probability distribution
(usually with an unknown normalizing prefactor)
in a high-dimensional configuration space.
One often has to resort
to using a Markov chain Monte Carlo method,
which converges only in the limit to the prescribed distribution.
Such methods typically inch through configuration space step by step,
with acceptance of a step based on a Metropolis(-Hastings) criterion.
An acceptance rate of 100\% is possible in principle by
embedding configuration space in a higher-dimensional phase space and using
ordinary differential equations.
In practice, numerical integrators must be used, lowering the acceptance rate.
This is the essence of {\em hybrid Monte Carlo} methods.
Presented is a general framework for constructing such methods
under relaxed conditions:
the only geometric property needed is (weakened) reversibility;
volume preservation is not needed.
The possibilities are illustrated by deriving
a couple of explicit hybrid Monte Carlo methods,
one based on barrier-lowering variable-metric dynamics
and another based on isokinetic dynamics.
\end{abstract}


\maketitle

\section{Summary}

\topic{importance}
Generating random samples from a prescribed distribution
is of immense importance
in the simulation of materials and in Bayesian statistics and machine learning.
For example, for a protein, just a single (unbiased) sample of its structure
can, with high probability, reveal its native structure.

\topic{Computational Task}
The computational task is to generate
samples from a distribution, whose probability density function
\[\rho_\rmx(x)\propto\exp(-V(x)),
\quad x = [x_1, x_2,\ldots, x_N]\T,\]
is known up to a multiplicative constant.
We assume that the state variables have been scaled and rendered dimensionless.
If $V(x)$ denotes energy or enthalpy, its value is in units of $k_\mathrm{B}T$.

\topic{MRRTT methods}
To generate unbiased samples generally requires a Markov chain Monte Carlo
(MCMC) method of the type proposed in the 1953 landmark paper by N.~Metropolis,
A.~Rosenbluth,  M.~Rosenbluth, A.~Teller, and E.~Teller (MRRTT),\cite{MRRT53}
which uses conditional acceptance to enforce unbiased sampling.
Unfortunately, such methods, if based on random walk proposals,
generate highly correlated samples,
so many steps are required to produce effectively independent samples.

\topic{HMC}
Hybrid Monte Carlo (HMC) methods,
introduced in 1987~\cite{DKPR87} and generalized in 1991,\cite{Horo91}
offer the possibility of reducing correlation between successive samples.
Such methods are based
on the numerical integration of ordinary differential equations (ODEs).
Key requirements in their theoretical justification~\cite{MeHF92,Sanz14} are
that the dynamics and its discretization must be reversible and
preserve phase-space volume.
Yet, there is in the literature%
\footnote{See Ref.~\cite{LeRe09}, p.~126 of Ref.~\cite{GiCa11}
 and Ref.~\cite{LSSG12}.}
a realization that the assumption of volume conservation
can be relaxed by including a Jacobian
in the probability ratio of the acceptance test.
Here we fully develop this idea
and also weaken the reversibility assumption,
providing a general framework for designing novel HMC dynamics
under relaxed conditions.

\topic{contributions}
The convergence of an MCMC method relies on two properties:
stationarity and ergodicity.
Presented in this article is a significant weakening
of sufficient conditions for stationarity:
preservation of volume in phase space is not required
and reversibility is required
only in the form of a bijection rather than an involution.
The omission of the volume preservation requirement has been noted
already for special cases.\cite{LeRe09,LSSG12}
Weaker requirements
expand possibilities for designing better proposals for MCMC moves.
In particular, many ``purely dynamical'' samplers can be made unbiased,
e.g., the non-Hamiltonian sampling dynamics presented in Ref.~\cite{TLCM01}.
The construction of a hybrid Monte Carlo method begins with the
selection of a system of ODEs in a higher-dimensional phase space
having an invariant probability density whose marginalization to the given
state variables $x$ is the target density $\rho_\mathrm{x}(x)$.
Three examples, two novel, are given for such a construction,
all of which violate phase-space volume preservation.
An HMC method requires a reversible numerical integrator,
but it can be difficult to find one that is efficient%
---unless a change of variables is effected.
Given in this article is a simple general formula for changing variables,
a special case occurring in Ref.~\cite{LSSG12}.
It is worth noting
that the analytical formulations and derivations presented here
require only elementary calculus
(and not differential geometry\cite{TLCM01,GiCa11}).

\topic{presentation style}
A familiarity with basic MCMC methods is assumed so that
the ideas can be presented in an abstract style that highlights key concepts.
The mathematics requires substantial use of the chain rule:
clarity is enhanced (and mistakes avoided!)
by liberal use of the function-composition operator~$\circ$,
e.g., $f\circ g(x)$ instead of $f(g(x))$.
To avoid excessive parentheses,
the raised dot is used for scalar and matrix multiplication.

\subsection{Discussion}

\topic{usefulnes of generalization}
The extent to which the relaxed requirements can aid in the creation
of better MCMC propagators remains to be seen.
Yet, there are already a couple of examples from the work of others.
Also included are some simple numerical experiments with results
that indicate potential benefit for the generalizations
developed in this article.

\topic{HMC vs. MCMC}
For molecular simulation, one of the advantages of traditional MCMC over MD
(or HMC) is the ability to use a set of {\em large} moves tailored
to the characteristics of the problems, e.g.,
concerted rotations of dihedral angles
that deform a polymer only locally.\cite{HuMD06}
It is conceivable that ODEs can be designed that make tailored moves.

\subsection{Outline of paper}

Section~2 of this article is a review of basic ideas of MCMC and
especially HMC, which also serves to introduce notation.
Section~3 presents the generalized MCMC and HMC methods.
Three examples follow in Section~4.
Discussion of (modified) detailed balance is deferred until Section~5.
It is more straightforward, it seems, to verify stationarity directly
than to use detailed balance as a stepping stone.
The importance of detailed balance is its role in assuring
self-adjointness of the operator that propagates probabilities.

\section{MCMC Methods}

This section reviews basic ideas.

\subsection{Markov chain methods}

\topic{aim}
Consider a Markov chain in configuration space
$\bfx^0\rightarrow \bfx^1\rightarrow\cdots\rightarrow \bfx^n$
that samples from a distribution with
unnormalized density $\exp(-V(x))$,
where boldface denotes a random variable.
The aim is to estimate ``observables''
$\langle A(\bfx)\rangle = \int A(x)\rho_\rmx(x)\rmd x$
by means of MCMC path averages
$\bar{\mathbf{A}}_n = (1/n)\sum_{k=1}^n A(\bfx^k)$.

\topic{MD vs. MC}
In molecular simulation, at least,
there are two ways to generate such Markov chains.
The first is to use discretized (stochastic) differential equations.
Such an approach produces a systematic error due to discretization,
and for this reason, it is shunned by the statistics community.
The second way to generate a Markov chain is by
using an MRRTT method,\cite{MRRT53} based on conditional acceptance.
This produces no systematic error (in the limit as $n\rightarrow\infty$).
Not surprisingly, such an approach tends to be less efficient,
and it does not scale as well with $N$.
It is a research challenge, e.g, Ref.~\cite{LeRe09},
to make such methods competitive with those
that unconditionally accept each move.

\topic{two properties}
Convergence for a Markov chain depends on two properties:
\begin{enumerate}
\item {\em stationarity}, which means that
 if $\bfx^k$ has probability density $\rho_\rmx(x)$,
 so will $\bfx^{k+1}$; and
\item {\em ergodicity}, which means that
 the chain almost surely visits every set of positive probability.
\end{enumerate}
Ergodicity is typically ensured by inclusion of stochastic randomness.

A {\em nonstandard} introduction to MRRTT methods follows,
which enables the seamless inclusion of hybrid Monte Carlo as a special case.

\subsection{Basic Monte Carlo methods}

Two examples are given of the process of generating a sample
$\bfx^{k+1}$ from $\bfx^k$.

\subsubsection{Random walk sampler}

Given a sample $\bfx$ be given from the target p.d.f.\ $\rho_\rmx(x)$,
the next sample is determined as follows:
\begin{enumerate}
\item[(1)] Generate auxiliary random variables $\bfy$
 from a Gaussian distribution with mean $\bfx$ and covariance matrix $2\tau I$.
 Hence,
 $\rho(x, y) = \rho_\rmx(x)(4\pi\tau)^{-N/2}\exp(-(y - x)^2/(4\tau))$
 serves as a joint probability density for $(\bfx, \bfy)$.
\item[(2a)] Construct a proposed move: $(\bfx', \bfy') = (\bfy, \bfx)$.
 With probability
 \[ \min\left\{1,\frac{\rho(\bfx',\bfy')}{\rho(\bfx, \bfy)}\right\}\]
 choose the next sample to be $\bfx'$.
\item[(2b)] Otherwise, choose it to be $\bfx$ again.
\end{enumerate}
Due to symmetry in the proposal,
the probability ratio simplifies to $\rho_\rmx(\bfx')/\rho_\rmx(\bfx)$.

\subsubsection{Brownian sampler}

\topic{the scheme}
This scheme aims at a higher acceptance rate by
generating $\bfy$ from a Gaussian distribution
with mean $\bfx -\tau\nabla V(\bfx)$ and covariance matrix $2\tau I$.
The limit $\tau\rightarrow 0$ yields Brownian dynamics.
The basic idea is similar to the force-bias MC method~\cite{PaRB78}
and almost identical to smart Monte Carlo,\cite{RoDF78} both proposed in 1978.
Extension to general densities is given in Refs.~\cite{Besa94,RoTw96}
and is named MALA in the latter reference.
The asymmetric proposal of this scheme
is an example of the Hastings~\cite{Hast70} generalization of the MRRTT
acceptance-test ratio.

\topic{transition}
Drift-diffusion dynamics is a slow way to explore configuration space.
Addition of inertia remedies this.
Moreover, there is a very broad set of possibilities based on approximating
the flow of a dynamical system
having an invariant density $\rho(x, y)$
whose marginalization to $\bfx$ is $\rho_\rmx(x)$
Indeed, the name ``hybrid'' stems from the use of {\em dynamics}
to constuct proposals.

\subsection{Hybrid Monte Carlo methods}

\topic{Hamiltonian Flow}
With a change of auxiliary variables $y = x +\sqrt{2\tau}p$
in the random walk sampler, the probability density becomes
(see formula~(\ref{eq:cvrho}))
\[\rho(x, p)
 =\rho_\rmx(x)(2\pi)^{-N/2}\exp(-p^2/2)\propto\rme^{-H(x, p)}\]
where $H(x, p) = V(x) +\frac12 p\T p$.
The value $p_i$ serves as the momentum for configuration variable $x_i$.
This density
is preserved by a Hamiltonian flow $\Phi_\tau$ of any duration $\tau$,
where the flow $\Phi_t$, $0\le t\le\tau$ is given by
\[\frac{\rmd}{\rmd t}\Phi_t(z) = f(\Phi_t(z)),\quad\Phi_0(z) = z,\]
with
\[z = \left[\begin{array}{c}x\\p\end{array}\right],\quad f(z) =
 \left[\begin{array}{c}\nabla_p H(x, p)\\-\nabla_x H(x, p)\end{array}\right].\]
To define a Hamiltonian system,
all you need is an even number of variables,
which can always be arranged by introducing additional variables.

\topic{Hybrid Monte Carlo}
Hybrid Monte Carlo is introduced in a seminal paper
by Duane, Kennedy, Pendleton and Roweth in 1987.~\cite{DKPR87}
The HMC propagator is $\Psi\approx\Phi_\tau$ where $\Psi$ is composed of
$\nu$ steps of a numerical integrator $\Psi_{\dt}\approx\Phi_{\dt}$
with $\dt =\tau/\nu$.
For example, the St\"ormer/Verlet/leapfrog scheme is given by
\begin{equation}  \label{eq:lfa}
 \Psi_\dt
 =\Phi^\mathrm{B}_{\dt/2}\circ\Phi^\mathrm{A}_{\dt}\circ\Phi^\mathrm{B}_{\dt/2}
\end{equation}
where
\begin{equation}  \label{eq:lfb}
 \Phi^\mathrm{A}_t(x, p) =
 \left[\begin{array}{c}x + t p \\ p\end{array}\right],\quad
 \Phi^\mathrm{B}_t(x, p) =
 \left[\begin{array}{c}x \\ p - t\nabla V(x)\end{array}\right].
\end{equation}
are each the exact flow of a piece of the Hamiltonian system.

\topic{algorithm}
Assuming a joint density $\rho(x,y)$ has been chosen
for which $\int\rho(x,y)\rmd y =\rho_\rmx(x)$,
a general recipe for HMC is the following:
Let $\bfx$ be a sample from the target density $\rho_\mathrm{x}(x)$.
\begin{enumerate}
\item[(1)] Generate $\bfy$ from the marginalized distribution with density
 $\rho(y|\bfx) =\rho(\bfx, y)/\int\rho(\bfx, y)\rmd y$
 and set $\bfz =[\begin{array}{cc}\bfx\T&\bfy\T\end{array}]\T$.
\item[(2a)] Calculate a proposed move:
 $\bfz' =\Psi(\bfz)$.
 With probability
\[ \min\left\{1,\frac{\rho(\bfz')}{\rho(\bfz)}\right\}\]
 choose the next sample to be $\bfx'$
 where $[\begin{array}{cc}(\bfx')\T&(\bfy')\T\end{array}]\T =\bfz'$.
\item[(2b)] Otherwise, choose it to be $\bfx$ again.
\end{enumerate}

\topic{sufficient conditions for stationarity}
For HMC to satisfy stationarity, it is sufficient that~\cite{MeHF92}
\begin{enumerate}
\item $\Psi$ be reversible w.r.t.\ $\rho$,
 meaning that there is a linear mapping $R$
such that $R\circ R = \mathrm{id}$, $\Psi^{-1}= R\circ\Psi\circ R$,
and $\rho\circ R = \rho$.
For classical HMC, $R(z) = [\begin{array}{cc}x\T -p\T\end{array}]\T$.
\item $\Psi$ be volume-preserving, meaning that $|\calJ\Psi| = 1$
 where \fbox{$\calJ\Psi =\det\partial_z\Psi$} is the Jacobian
(determinant of the Jacobian matrix).
\end{enumerate}

\topic{HMC integrators}
Although the leapfrog method is only second order accurate,
its accuracy is difficult to beat for larger step sizes.
An alternative is proposed in Ref.~\cite{BlCS1y},
which tunes a second order splitting to a target density that is Gaussian.
Another alternative is shadow Hamiltonian HMC,\cite{IzHa04,SHSI09,AkRe12}
which samples from a modified Hamiltonian that is better conserved
by the leapfrog integrator,
but it requires a slight reweighting of the samples.
An important practical issue is that of choosing the step size.
See Ref.~\cite{BPRS13} for a theoretical study with references to earlier work

\subsection{Generalized MCMC}

\topic{motivation}
In 1991, Horowitz~\cite{Horo91} generalizes HMC by doing MCMC in phase space,
\[\bfz^0\rightarrow \bfz^1\rightarrow\cdots\rightarrow \bfz^n,\]
with only a partial refresh of the auxilliary variables
from one step to the next.
This modification allows samples to be taken more frequently:
one can use a smaller duration $\tau$ without losing momentum.

\topic{algorithm}
Given a sample $\bfx$, $\bfy$ from the chosen joint distribution,
the next sample is determined as follows:
\begin{enumerate}
\item[(1)] Make a random change $\bfy'$ to $\bfy$
 that preserves density $\rho(\bfx, y)$.
\item[(2a)] Calculate a proposed move $\bfz'' =\Psi(\bfz')$.
 With probability
\[ \min\left\{1,\frac{\rho(\bfz'')}{\rho(\bfz')}\right\}\]
 choose the next sample to be $\bfz''$.
\item[(2b)] Otherwise, choose it to be $R(\bfz')$.
\end{enumerate}
The reversal \substep~(2b) is, of course, an unfortunate outcome,
because it means going backwards on the same trajectory
and only slowly diffusing away from it.

\topic{range of possibilities}
There are a range of possibilities for \substep~(1).
At one extreme, regular HMC generates a fresh $\bfy'$ in each step;
at the other extreme with $\bfy' =\bfy$,
the result is a random walk along a single discrete trajectory
(which is almost certain not to be ergodic,
and, even if it were, convergence would be extremely slow).

\subsubsection{Langevin sampler}

A specific example is the method L2MC of Horowitz:\cite{Horo91}
Given a sample $\bfx,\bfp$,
it generates $\bfp'$ from a Gaussian distribution with
mean $\sqrt{1 - 2\gamma\tau}\bfp$ and covariance matrix $2\gamma\tau I$.
A configuration space proposal is obtained from
$\bfz'' =\Psi_\tau(\bfz')$ where $\Psi_\tau$ is the leapfrog integrator
defined by Eqs.~(\ref{eq:lfa}) and~(\ref{eq:lfb}).
The limit $\tau\rightarrow 0$ gives Langevin dynamics.

\subsection{Variable-Metric Hybrid Monte Carlo}

The integration step size $\dt$ is limited by the fastest time scale.
Inserting a metric tensor into the ``kinetic energy'' term
can compress the range of time scales, resulting in faster sampling
within a basin of the potential energy surface.%
\footnote{See Sec.~3.4 of Ref.~\cite{BPSS11}.}
The idea is to integrate a Hamiltonian system with Hamiltonian
\[H(x, p) = V(x) +\frac12 p\T M(x)^{-1}p +\frac12\log\det M(x).\]
This has invariant density $\propto\exp(-H(x, p))$, and
the  marginal density for $x$ is $\propto\exp(-V(x))$.
With $M_k = (\partial/\partial x_k)M$ and with $e_k$ denoting the unit vector
for the $k$th coordinate direction, one obtains the equations
\begin{eqnarray}
\frac{\rmd}{\rmd t}x &=& M^{-1}p, \nonumber \\
\frac{\rmd}{\rmd t}p &=& -\nabla_x V
 -\frac12\sum_k(\mathrm{tr}(M^{-1}M_k) - p\T M^{-1}M_k M^{-1}p)e_k,
  \label{eq:vmhmc}
\end{eqnarray}
where $\mathrm{tr}$ denotes the trace.

The idea of mass-tensor dynamics to improve sampling goes
back as far as Bennett in 1975.\cite{Benn75}
That paper mentions but does not develop the idea of nonconstant mass
tensors for sampling.
It is, of course, unaware of the importance of using a geometric integrator.
The special case of a constant diagonal mass matrix
is explored for molecular simulation in Ref.~\cite{LiTu10}.

Recently,
Girolami and Calderhead~\cite{GiCa11} propose the use of mass-tensor dynamics
with a symplectic integrator for hybrid Monte Carlo for Bayesian statistics.
For the mass matrix, they
propose the use of the expected Fisher information matrix
plus the negative Hessian of the log-prior.
For Bayesian inference,
the expected Fisher information matrix is
normally positive definite and always semi-positive definite.
A numerical experiment%
\footnote{See Fig.~1 in Ref.~\cite{LSSG12}.}
for a banana-shaped distribution
illustrates the potential effectiveness of this approach
for Bayesian statistics.

\section{General Theory}

Presented here are the most general conditions
known to be sufficient for stationarity.
Previously stated results are special cases.

\subsection{General MRRTT Methods}

\topic{two ingredients}
The target density $\rho_\rmx(x)$ is given, but the phase-space density
$\rho(x, y)$ and the proposal mapping $\Psi(x, y)$ must be chosen.
It is not assumed that $y$ has the same dimension as $x$.

\topic{joint density}
It is required that $\int\rho(x,y)\rmd y =\rho_\rmx(x)$.
As a practical matter, it must be easy to sample $\bfy$ from
the conditional density $\propto\rho(\bfx, y)$.

\topic{mapping}
In practice,
$\Psi(x, y)$ may be formally undefined for some combinations of $y$ with $x$,
typically
because the resulting $x'$ is outside (the designated) configuration space.
Let $\Omega$ be that part of phase space on which $\Psi$ is defined.
The one requirement on $\Psi$ is that of {\em generalized reversibility}:
$\Psi$ is said to be {\em reversible} w.r.t.\ $\rho$ if
\begin{equation}  \label{eq:grtwo}
\Psi^{-1} = R^{-1}\circ\Psi\circ R
\end{equation}
where $R$ is a bijection on phase space that preserves probability, i.e.,
\begin{equation}  \label{eq:grone}
\rho\circ R = \rho/|\calJ R|,
\end{equation}
and is such that $R\circ\Psi$ is a bijection on $\Omega$.
Note that $|\calJ R| = 1$ if both $R$ is linear and $R\circ R = \mathrm{id}$.

The generalization to compressible proposal mappings
is enabled by including the factor $|\calJ\Psi(\bfz')|$
in the probability ratio of the acceptance test:

\begin{algorithm}  \label{alg:one}
\caption{Compressible Generalized Monte Carlo}
Let $\bfz = [\begin{array}{cc}\bfx\T &\bfy\T\end{array}]\T$ be given.
One step of the Markov chain is as follows:

\qquad(1) Generate $\bfy'$
from the marginalized distribution with density
 $\rho(y'|\bfx) =\rho(\bfx, y')/\int\rho(\bfx, y)\rmd y$
 and set $\bfz' = [\begin{array}{cc}\bfx\T &(\bfy')\T\end{array}]\T$.

\qquad(2a) If $\bfz'\in\Omega$,
 set $\bfz'' =\Psi(\bfz')$ and
 accept $\bfz''$ with probability
\[
\min\left\{1, \frac{\rho(\bfz'')}{\rho(\bfz')}|\calJ\Psi(\bfz')|\right\}.
\]
\qquad

\qquad(2b) If $\bfz'\not\in\Omega$ or $\bfz''$ is rejected,
choose $R(\bfz')$.
\end{algorithm}

{\it Notes.}
(i) It is {\em not} required that $\bfz\in\Omega$,
because \substep~(1) can produce a value $\bfz'\in\Omega$.
(ii) The specified density may contain Dirac delta functions.
These represent conservation laws,
and the propagator must enforce them exactly.

It is shown in Appendix~\ref{sss:stat}  
that the second \substep\ of phase space HMC satisfies stationarity
if the propagator is reversible.
A direct demonstration of stationarity is
easier than proving detailed balance,
which is the topic of Sec.~\ref{sec:detail}.

\subsection{Flows}  \label{ss:dflow}

\topic{continuity equation}
The benefit of compressibility for HMC
is that the underlying continuous flow $\Phi_\tau$ can be non-Hamiltonian.
The dynamics must nonetheless preserve the chosen joint density,
which we express as
$\rho(x, y) = \rho(\zee)\propto\exp(-H(\zee))$,
where $H(\zee)$
does not necessarily play the role of
a Hamiltonian.
For a flow $\Phi_t$ to have $\rho$ as an invariant density, it must hold that
\[\int_{\Phi_t(A)} \rho(z)\rmd z =\int_A \rho(z)\rmd z\]
for an arbitrary set $A$.
Suppose the flow $\Phi_t$ be defined in terms of a vector field $f(\zee)$
by an ODE system $(\rmd/\rmd t)\Phi_t = f\circ\Phi_t$, $\Phi_0 =\mathrm{id}$.
It has invariant density $\rho$
if the continuity equation $\nabla\cdot(\rho f) = 0$
is satisfied.\cite{Skee09a}
This is conveniently expressed in terms of $H(z)$ by
\begin{equation} \label{eq:feq}
f\cdot\nabla H = \nabla\cdot f.
\end{equation}

\topic{reversibility}
From Eqs.~(\ref{eq:grtwo}) and~(\ref{eq:grone}), it follows that
the flow $\Phi_t$ is reversible w.r.t.\ $\rho$ if and only if
there is a bijection $R$ such that
\begin{equation}  \label{eq:rtwo}
f\circ R = -(\partial_z R)f,
\end{equation}
and
\begin{equation}  \label{eq:rone}
H\circ R = H + \log|\calJ R|.
\end{equation}
To obtain Eq.~(\ref{eq:rtwo}),
apply $\rmd/\rmd t$ to
$\Phi_t\circ R = R\circ\Phi_{-t}$,
which follows from Eq.~(\ref{eq:grtwo}),
and set $t = 0$.
If $R$ is linear, then $(\partial_\zee R)f$ simplifies to $R\circ f$.

\topic{the design process}
Therefore, once
$\rho(\zee)$, and therefore $H(\zee) = -\log\rho(\zee) + \mathrm{const}$,
has been chosen,
one needs to find $f(z)$ and $R(z)$ that satisfy
continuity equation~(\ref{eq:feq}) and
reversibility conditions~(\ref{eq:rone})/(\ref{eq:rtwo}).
The aim is to design an ODE $(\rmd/\rmd t)z = f(z)$ which
explores state space rapidly and
accommodates an {\em explicit} reversible integrator
with large step sizes $\dt$.
This is a different perspective from that of Ref.~\cite{TLCM01},
where the system of ODEs rather than the invariant density
is assumed to be given.

\subsection{Integrators from splittings}

\topic{splitting methods}
We consider the use of splitting methods to approximate the dynamics.
Splitting methods are easy to design both to be reversible and
to be explicit with respect to force $F(x) = -\nabla V(x)$ and energy $V(x)$.
The idea is to express $f$ as a sum of terms $f^\kappa$
such that each system $(\rmd/\rmd t)\zee = f^\kappa(\zee)$
can be integrated analytically (or numerically using a reversible
discretization that is explicit with respect to force and energy).
A symmetric splitting gives at least second order accuracy.

\topic{an example}
For the examples that follow, assume a Trotter-Strang splitting
\[f = \frac12\fb + \fa + \frac12\fb\]
with the proposal calculated using
\[\zee_{1/4} =\Phib_{\dt/2}(\zee_0),\quad
  \zee_{3/4} =\Phia_\dt(\zee_{1/4}),\quad
  \zee_1 = \Phib_{\dt/2}(\zee_{3/4}).\]
Then the acceptance test ratio would be a product of factors of the form
\[\frac{\rho(\Psi_\dt(\zee_0))}{\rho(\zee_0)}\calJ\Psi_\dt(\zee_0)
 =\exp(H(\zee_0) - H(\zee_1))\calJ\Phib_{\dt/2}(\zee_0)
 \calJ\Phia_\dt(\zee_{1/4})
 \calJ\Phib_{\dt/2}(\zee_{3/4}).\]

\topic{conservation laws}
If the density $\rho(z)$ contains Dirac delta function(s),
the associated conservation law must, in practice,
be satisfied by each part of the splitting.
Alternatively,
the conservation law might be used to eliminate a degree of freedom.
An example of each possibility
is given in Sections~\ref{ss:iso} and~\ref{ss:nh}, respectively.

\topic{compressibility integral}
If $(\rmd/\rmd t)\zee = f^\kappa(\zee)$ is integrated analytically,
the exact value of $\calJ\Phi^\kappa_t(\zee)$ is needed.
Its formula can be obtained directly from the formula for $\Phi^\kappa_t$.
Alternatively, for some vector fields $f^\kappa(\zee)$,
one can find a {\em compressibility integral}\/~\cite{Skee09a}
$\ci^\kappa(\zee)$
such that $f^\kappa\cdot\nabla\ci^\kappa = \nabla\cdot f^\kappa$,
in which case
one can use
\begin{equation}  \label{eq:ci}
 \calJ\Phi^\kappa_t =\exp(\ci^\kappa\circ\Phi^\kappa_t - \ci^\kappa).
\end{equation}
This is derived in Sec.~\ref{sss:jac}, which follows.

\subsubsection{Derivation of a formula for Jacobian of a flow}  \label{sss:jac}

Omiting the superscript $\kappa$, we have
\[\frac{\rmd}{\rmd t}\det\partial_\zee\Phi_t
 = (\det\partial_\zee\Phi_t)
  \mathrm{tr}((\partial_\zee\Phi_t)^{-1}\frac{\rmd}{\rmd t}\partial_\zee\Phi_t)
 = (\det\partial_\zee\Phi_t)\mathrm{tr}(\partial_\zee f\circ\Phi_t) \]
where the first step
uses the formula for the derivative of a determinant and the second step
results from interchanging $\rmd/\rmd t$ with $\partial_\zee$.
Hence,
\[\frac{\rmd}{\rmd t}\log|\det\partial_\zee\Phi_t|
 = (\nabla\cdot f)\circ\Phi_t,
\quad\mbox{and}\quad
\calJ\Phi_t = \exp\int_0^t(\nabla\cdot f)\circ\Phi_s\,\rmd s.\]
The result follows from
$(\nabla\cdot f)\circ\Phi_t = (f\cdot\nabla\ci)\circ\Phi_t
 = (\rmd/\rmd t)\ci\circ\Phi_t$.

\subsection{Change of variables}  \label{ss:chvar}

A change of variables can facilitate numerical integration.
This is illustrated in Sec.~\ref{ss:vm}
with an example from Eq.~(8) of Ref.~\cite{LSSG12}.

Assume Eqs.~(\ref{eq:feq})--(\ref{eq:rtwo}) hold.
Let $z =\zeta(w)$ be a bijection from some set $\bar{\Omega}$ to $\Omega$.
Then the following relationships hold:
The random variable $\bfw =\zeta^{-1}(\bfz)$ has p.d.f.\ 
$\bar{\rho}(w)\propto\exp(-\bar{H}(w))$ where
\begin{equation}  \label{eq:cvrho}
\bar{H} = H\circ\zeta -\log|\det\partial_w\zeta|.
\end{equation}
The transformed map
\begin{equation}  \label{eq:cvphi}
\bar{\Phi}_t = \zeta^{-1}\circ\Phi_t\circ\zeta\quad
\mbox{has invariant density}\quad\bar{\rho}(w).
\end{equation}
The map $\bar{\Phi}_t$
satisfies $(\rmd/\rmd t)\bar{\Phi}_t = \bar{f}\circ\bar{\Phi}_t$
for the vector field $\bar{f}(w)$ given by
\begin{equation}  \label{eq:cvf}
 \bar{f} = (\partial_w\zeta)^{-1} f\circ\zeta.
\end{equation}
The flow $\bar{\Phi}_\tau$ is reversible w.r.t.\ $\bar{\rho}$
for the bijection $\bar{R}(w)$ given by
\begin{equation}  \label{eq:cvr}
 \bar{R} = \zeta^{-1}\circ R\circ\zeta.
\end{equation}
Derivations for Eqs.~(\ref{eq:cvrho})--(\ref{eq:cvr})
are provided in Appendix~\ref{sss:cv}.

\section{Three Examples}

For the first two of these examples, it seems that viable algorithms
are possible only for compressible formulations of the dynamics.

\subsection{Explicit Variable-Metric HMC}  \label{ss:vm}

The Hamiltonian variable-metric HMC method
requires solving nonlinear equations involving a variable metric tensor $M(x)$
at each time step.
Motivated by Lagrangian mechanics, it is shown in Ref.~\cite{LSSG12}
that solving equations can be avoided using compressible HMC.

Applying the change of variables formula of Sec.~\ref{ss:chvar}
for $(x, p) = \zeta(x, v) = (x, M(x)v)$, one obtains
\[\bar{H}(x, v) = V(x) +\frac12 v\T M(x)v -\frac12\log\det M(x).\]
From
\[\partial_w\zeta
 = \left[\begin{array}{cc} I & 0 \\ \sum_k M_k v e_k\T & M \end{array}\right]\]
and Eq.~(\ref{eq:vmhmc}), one obtains the equations
\begin{eqnarray}
\frac{\rmd}{\rmd t}x &=& v, \nonumber \\
M\frac{\rmd}{\rmd t}v &=& -\nabla_x V
 -\frac12\sum_k(\mathrm{tr}(M^{-1}M_k) - v\T M_k v)e_k
 -\sum_k v_k M_k v,  \label{eq:dvdt}
\end{eqnarray}
which agrees with Eq.~(8) of Ref.~\cite{LSSG12}.

With $(\rmd/\rmd t)x = 0$ in the equations for part B of the splitting,
this leaves quadratic differential equations to solve for $v$.
These are discretized~\cite{LSSG12} using a symmetric linearly implicit
scheme of Kahan (1993),\cite{Sanz94,KaLi97}
based on geometric averaging of terms quadratic in $v$.
This requires solving one linear system of equations per time step,
instead of iteratively solving a nonlinear system at each time step.

\subsubsection{Barrier-lowering dynamics}  \label{ss:barr}  

The effect of decreasing mass in a small region $X$ of
configuration space is to increase speed across that region.
Since the probability of being in $X$ is unchanged,
the region $X$ must be crossed more often.
This principle can be used to increase barrier crossings.
Let $x^0$ be the approximate location of a saddle point
and $m$, $m\T m = 1$, the direction of negative curvature.
Specifically, assume the Hessian of $V(x)$ at $x = x^0$
has a single negative eigenvalue with corresponding eigenvector $m$.
Define
\[ M(x) = \mu(x)^{N-1}m m\T + \mu(x)^{-1}(I - mm\T) \]
where
\[\mu(x) = 1 - (1 -\mu_0)\exp(-\frac{\|x - x^0\|^2}{2 r^2})\]
localizes the effect to a neighborhood of $x^0$.
Here $r$ is perhaps the distance to an inflection point
and $\log(1/\mu_0^2)$ is some fraction of the barrier height.
The ``tunnel metric'' proposed independently in Ref.~\cite{LaSS13} is similar.
However, it uses a switching function that is different in form
and defined by two points rather than just $x^0$.
And it lacks the normalization $\det M(x) = 1$,
which helps to reduce integration error.

To handle several such barriers, an elegant approach is to use
\[ M = \exp(\sum_\kappa\log M^\kappa),\]
which, in practice,
would be replaced by a Trotter or Trotter-Strang splitting.

\begin{figure}
\eps{\includegraphics[width=10cm]{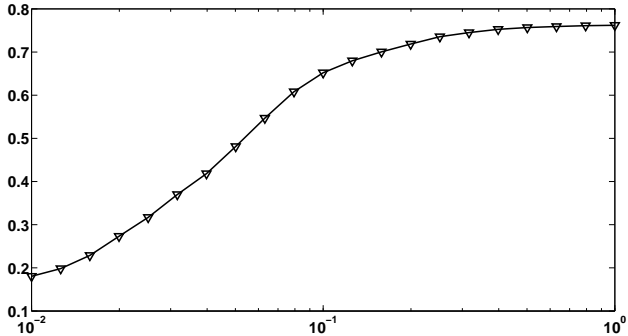}%
\caption{\label{fig:acc}
x-axis is the minimum mass and y-axis is the acceptance probability.}
} 
\end{figure}

\begin{figure}
\eps{\includegraphics[width=10cm]{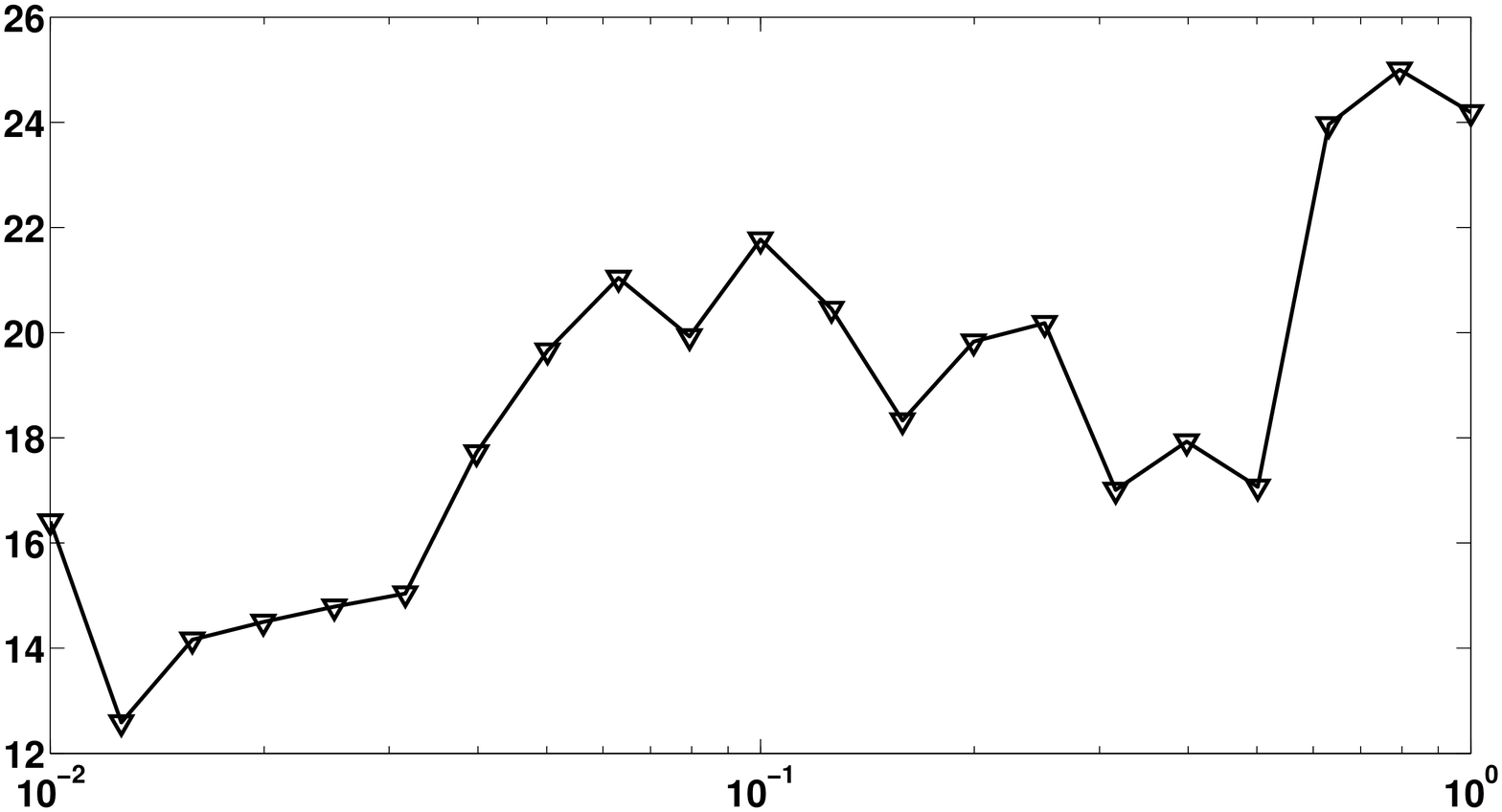}%
\caption{\label{fig:auto}
x-axis is the minimum mass and y-axis is the auto-correlation time.}
}
\end{figure}

The method is tested on a high-dimensional mixture of two Gaussians.
Specifically, the first dimension is the average of two Gaussians
located at $\pm 2.5$, and the other $128$ dimensions are all Gaussians.
Hence, the components of $\bfx$ are independent.
The standard deviation of each Gaussian in the first dimension is 1,
and the standard deviations in other dimensions
are uniformly distributed from 1 to 2.
This is the simplest problem
that features high dimensionality and multiple modes
and is not especially sensitive to the duration $\tau$ of the dynamics run.
(And it can be implemented very efficiently.)
The observable is a sigmoid function of the first dimension,
$A(x) = 1/(1 +\exp(-x_1))$,
which is chosen
to mimic the second eigenfunction shown in Fig.~1 of Ref.~\cite{ScHu00}.
The step size $\dt=0.5$, the number of integration steps $\nu=10$,
and the number of samples $n=10^6$.
This many samples gives an estimated standard deviation of 3
for the mean value of
the auto-correlation time
for HMC with an MC step of duration $\tau = 5$
and a number of integrator steps $\nu = 10$.
The auto-correlation time, computed by {\it Acor},\cite{Good09}
gives the ratio of the sample size to the {\em effective sample size}.
Results are displayed in Figures~\ref{fig:acc} and~\ref{fig:auto}
for $m$ in the direction of the x$_1$-axis, $x_0 = 0$, $r = 1$,
and a range of values of the minimum mass $\mu_0^{N-1}$.
For a fairly wide range of values of $\mu_0^{N-1}$,
the barrier lowering variable mass method
produces about 100\% more effective samples.

\subsection{Isokinetic HMC}  \label{ss:iso}

\topic{equations and density}
The equations for isokinetic dynamics are
\[
 \frac{\rmd}{\rmd t}x =\frac{N-1}{N}p,\quad
 \frac{\rmd}{\rmd t}p =\left(I - \frac{p p\T}{p\T p}\right)F.
\]
where $F = -\nabla_x V$.
With masses thus chosen, the density
\[\rho(x, p)\propto\exp(-N V(x)/(p\T p))\delta(p\T p - N)\]
is invariant
and $\langle \bfp_i^2\rangle = 1$.
For the purpose of satisfying Eq.~(\ref{eq:feq}), $\delta$ may be taken
to be a mollified delta function.%
\footnote{In fact, $\delta$ can be taken to be any differentiable
function for which $\rho(x, p)$ is a valid p.d.f.}
Isokinetic dynamics
is intended for sampling from the canonical ensemble $\rme^{-V(x)}$.%
\cite{EvMo83}
There is current interest in isokinetic dynamics because it is less sensitive
to resonances induced by multiple time-stepping.\cite{MiMT03,OmKo13}

\topic{splitting}
A convenient splitting is the following:
\[
\fa = \left[\begin{array}{c} ((N-1)/N)p \\ 0 \end{array}\right],\quad
\fb = \left[
\begin{array}{c} 0 \\ F - ((p\T F)/(p\T p))p \end{array}
 \right].\]
To satisfy stationarity, $\Psi_{\dt}$ must conserve kinetic energy exactly,
which is true if the flow $\Phi^\mathrm{B}_t$ is obtained analytically.

\topic{analytical solution}
To solve the equations $(\rmd/\rmd t)\zee = \fb(\zee)$, introduce
scalar quantities $\xi = (F\T F)^{1/2}$, $\zeta = (p\T p)^{1/2}$,
and $\eta = F\T p/(\xi\zeta)$.
The equations then reduce to solving
\[ \frac{\rmd}{\rmd t}\eta = (\xi/\zeta)(1 -\eta^2)\]
where $\xi$ and $\zeta$ are constant.
The solution is
\[\eta(t) =\frac{\eta(0)\cosh(\xi t/\zeta) +\sinh(\xi t/\zeta)}%
                {\cosh(\xi t/\zeta) +\eta(0)\sinh(\xi t/\zeta)},\]
which can be written
\[\frac{\xi}{\zeta}\eta(t)
 = \frac1{\sigma(t)}\frac{\rmd\sigma}{\rmd t}(t)\quad
\mbox{where }\sigma(t) = \cosh(\xi t/\zeta) +\eta(0)\sinh(\xi t/\zeta).\]
From this, it can be shown that
\[p(t) =\frac1{\sigma(t)}\left(p(0) + \frac{\xi^2}{\zeta^2}
\left(\frac{\rmd\sigma}{\rmd t}(t) -\frac{\rmd\sigma}{\rmd t}(0)\right)
F\right),\]
which is equivalent to the formula given in Sec.~IV.B of Ref.~\cite{MiMT03}.

\topic{compressibility}
The compressibility $\nabla\cdot\fa = 0$, but
$\nabla\cdot\fb = -(N-1)\xi\eta/\zeta$.
The compressibility $\nabla\cdot\fb$ has an integral
$\ci^\mathrm{B} = \frac12(N-1)\log(1 -\eta^2)$;
whence, using Eq.~(\ref{eq:ci}), we have
$\calJ\Phi^\mathrm{B}_t(z(0)) = ((1 -\eta(t)^2)/(1 -\eta(0)^2))^{(N-1)/2}
 = (\sigma(0)/\sigma(t))^{N-1}$.

\topic{divergence-free form}
Isokinetic dynamics can be transformed to make it divergence-free,
see Ref.~\cite{Skee09a}.  
However, this involves placing the prefactor
$\exp(-(N-1)V(x)/(p\T p))$ into the equations of motion,
for which there seems to be no reversible integrator explicit in force
and energy.

\subsubsection{Numerical experiments}

Using the same test problem as in Section~\ref{ss:barr},
Hamiltonian HMC is compared to isokinetic HMC,
in both cases with unit masses.
Given in Tables~\ref{tab:ham} and~\ref{tab:iso}
are the estimated effective sample sizes per 1000 force evaluations for each,
for varying values of the duration $\tau$ of an MC step and
the number of integrator steps $\nu$ per MC step.
A pair of parameter values
that maximizes the number of effective samples per integrator step is
in the center of each table.
A dash indicates a failure of all proposed moves.
The isokinetic method performs a little better than the Hamiltonian method
and {\em is less sensitive to tuning parameters.}

\begin{table}
\caption{\label{tab:ham}
Effective number of samples per 1000 force evaluations for Hamiltonian HMC}
  \begin{tabular}{| l | c | c | c | c |}
  
    \hline
      $\tau$ $\backslash$ $\nu$ & 6 & 8 & 10 & 12 \\ \hline
    $4$ & 1.44 & 2.51 &3.03  & 2.55\\ \hline
    $5$ & 2.04 & 4.41 & 4.13 & 3.65\\ \hline
    $6$ & $-$ & 1.52 &3.42  & 3.19\\ \hline

  \end{tabular}
\end{table}

\begin{table}
\caption{\label{tab:iso}
Effective number of samples per 1000 force evaluations for isokinetic HMC}
  \begin{tabular}{| l | c | c | c | c |}
  
    \hline
      $\tau$ $\backslash$ $\nu$ & 6 & 8 & 10 & 12 \\ \hline
    $4$ & 3.16 & 3.52  &2.81  & 3.20\\ \hline
    $5$ & 3.83 & 4.52  & 4.91 & 4.84\\ \hline
    $6$ & 0.72 & 4.11 & 3.60 & 3.29\\ \hline

  \end{tabular}
\end{table}

\subsection{Nos\'e-Hoover thermostat}  \label{ss:nh}

The Nos\'e-Hoover thermostat employs the extended system
\[
 \frac{\rmd}{\rmd t}x = p,\quad
 \frac{\rmd}{\rmd t}p = F -\frac{\ps}{Q}p,\quad
 \frac{\rmd}{\rmd t}s = \frac{\ps}{Q},\quad
 \frac{\rmd}{\rmd t}\ps = p\T p - N.
\]
The invariant density is
\[\rho(\zee, s)\propto\rme^{Ns}\delta(H(\zee) + N s)\]
where $H(\zee) = \frac12 p\T p + V(x) +\ps^2/(2Q) - E$,
$Q$ is ``thermal mass'', and $E$ is some constant.
The vector $\zee$ consists of $x$, $p$, and a scalar $\pi$.
Again, time is normalized so that $\langle \bfp_i^2\rangle = 1$.

The quantity $H(\zee) + N s$ is conserved.
However, there seems to be no practical numerical integrator
that conserves this quantity exactly.
Therefore, we eliminate
the equation for $(\rmd/\rmd t)s$,
which is possible since no other variable depends on it,
to get a system with invariant density $\rho(\zee)\propto\exp(-H(\zee))$.

Leimkuhler and Reich~\cite{LeRe09}
develop a compressible generalized HMC version of this scheme,
in which only the thermostat variable $\pi$ is partially refreshed,
and they prove stationarity.
Their aim is to show the feasibility of introducing conditional acceptance
into stochastic sampling dynamics.
For an integrator,
they use a compact splitting of Ref.~\cite{Klei98}:
\[
\fa = \left[\begin{array}{c} p \\ 0 \\ p\T p - N\end{array}\right],\quad
\fb = \left[
\begin{array}{c} 0 \\ F -\ps p/Q \\ 0\end{array}
 \right]\]
(p.~746 of Ref.~\cite{LeRe09}).
The system $(\rmd/\rmd t)\zee = \fb(\zee)$
has nonzero compressibility $\nabla\cdot\fb = -N\ps$.
Obtaining the Jacobian directly gives
$\calJ\Phi^\mathrm{B}_t =\exp(-N\ps t/Q)$,
(from which one can obtain the compressibility integral
$\ci^\mathrm{B} = N\log\|F -\ps p/Q\|$).

There are a couple of ways of making the Nos\'e-Hoover scheme divergence-free,
see~Refs.\cite{TLCM01,Skee09a,LeRe09},
but in either case we are left with a system having a conservation law
whose exact enforcement seems impractical.

\section{Modified Detailed Balance}  \label{sec:detail}

\topic{(modified) detailed balance}
Many compressible MCMC schemes satisfy not only stationarity but also
a modified detailed balance property.
The benefit of this is that the operator that propagates probabilities
is self-adjoint, making it easier to analyze and understand
the behavior of the Markov chain.
A Markov process satisfies {\em (modified) detailed balance} if
\begin{equation} \label{eq:gdetbal}
 \rho(\zee'|R(\zee))\rho(\zee) =\rho(\zee|R(\zee'))\rho(\zee')
\end{equation}
where $\rho(\zee'|\zee)$ is the conditional p.d.f.\ for a step.
(The definition given in Ref.~\cite{Gard04} is recovered by replacing
$z$ by $R(z)$ and making the assumptions
$R\circ R = \mathrm{id}$ and $\rho\circ R =\rho$.)

\topic{stationarity}
It is easy to show that modified detailed balance implies stationarity:
Replace $z$ by $R(z)$ and make use of Eq.~(\ref{eq:grone}) to get
\[\int\rho(\zee'|\zee)\rho(\zee)\rmd\zee =
 \int\rho(\zee'|R(\zee))\rho(\zee)\rmd\zee.\]
The result follows from applying Eq,~(\ref{eq:gdetbal}) and using the fact that
\[\int\rho(\zee|R(\zee'))\rmd\zee = 1.\]

\subsection{The forward transfer operator and self-adjointness}

\topic{forward transfer operator}
The forward transfer operator (or propagator) $\calP$ relates
the {\em relative p.d.f.} $u_n(\zee) =\rho_n(\zee)/\rho(\zee)$
of a sample $\mathbf{\zee}_n$ to that of its predecessor $\mathbf{\zee}_{n-1}$:
\[\rho_n/\rho = u_n =\calP u_{n-1} =\calP(\rho_{n-1}/\rho).\]
Hence,
\[ \calP u(\zee) = \frac1{\rho(\zee)}
 \int\rho(\zee |\zee')u(\zee')\rho(\zee')\rmd\zee'\]
where $\rho(\zee |\zee')$ is the transition probability density.

\topic{self-adjointness}
It happens that the forward transfer operator $\calP$
is {\em self-adjoint} with respect to the $\rho$-weighted inner product
\[(u, v) =\int u(\zee)v(\zee)\rho(\zee)\rmd\zee,\]
defined on the space of functions for which $u(R(\zee)) =u(\zee)$
if the Markov process satisfies modified detailed balance.
We have
\[(\calP u, v) =\int\!\int\rho(z|z')u(z')\rho(z')v(z)\rmd z'\rmd z.\]
If we replace $z'$ by $R(z')$ and make use of Eq.~(\ref{eq:grone}), we get
\[(\calP u, v) =\int\!\int\rho(z|R(z'))u(z')\rho(z')v(z)\rmd z'\rmd z.\]
Similarly, we have
\[(\calP v, u) =\int\!\int\rho(z'|R(z))v(z)\rho(z)u(z')\rmd z'\rmd z\]
where we have interchanged not only $u$ and $v$
but also the names of the dummy variables.
Equality, $(\calP u, v) = (u,\calP v)$, follows from Eq.~(\ref{eq:gdetbal}).

\subsection{GHMC and detailed balance}

\topic{counterexample}
{\em Generalized HMC does not formally satisfy detailed balance.}
To illustrate why this is so, consider the
case of a Langevin sampler
for $V(x) = x^2/2$, $2\gamma\tau = 1$, $\tau =\pi/2$,
but with an exact integrator:
\begin{itemize}
\item[] Given scalars $x$ and $p$,
\item[(1)] replace $p$ with a sample $p'$ from a Gaussian distribution, and
\item[(2)] set $x'' = x + p'$ and $p'' = p' - x$.
\end{itemize}
The first \substep\ is a random move in the direction of the p axis;
the second is a 90$^\circ$ clockwise rotation in the x-p plane.
For most pairs $(x'', p'')$ generated this way,
it is {\em impossible} to get back to $(x, p)$ in a single MCMC step.

\topic{methods that satisfy detailed balance}
Generally, to ensure detailed balance for generalized HMC,
there should be a third \substep\ that
is the {\em time reversal}\/\cite{Croo00}
of the process given by the first \substep.
The reversal of a process with conditional p.d.f.\ $\rho(z'|z)$
has a conditional p.d.f.\ $\hat{\rho}(z'|z) = \rho(z|z')\rho(z')/\rho(z)$.
Specifically, it is verified in Appendix~\ref{sss:mdetbg}
that a complete step of a 3-\substep\ MCMC scheme
satisfies modified detailed balance if
\begin{enumerate}
\item The \substep~1 process has a conditional p.d.f.\ $\rho_1(z'|z)$
 that satisfies
\begin{equation}  \label{eq:cpdf}
\rho_1(R(z')|R(z)) =\rho_1(z'|z)/|\calJ R(z')|.
\end{equation}
\item The middle \substep\ satisfies detailed balance.
\item The \substep~3 process has a conditional p.d.f.
\[\rho_3(z'|z) = \rho_1(z|z')\rho(z')/\rho(z).\]
\end{enumerate}

\topic{detailed balance for GHMC proposals}
To ensure that a
{\em GHMC proposal \substep\ }
satisfies detailed balance,
make the additional assumption that $R$ is an involution:
\[ R\circ R = \mathrm{id}.\]
Under this assumption, it is shown in Appendix~\ref{sss:mdetb}  
that the second \substep\ of phase space HMC
satisfies modified detailed balance.

\topic{methods that effectively satisfy detailed balance}
The Langevin sampler can be made to formally satisfy detailed balance
by splitting the initial \substep\ into initial and final \substep s,
each of which multiplies the momenta by $(1 - 2\gamma\tau)^{1/4}$ and adds
independent Gaussians of mean 0 and variance $(1 - 2\gamma\tau)^{1/2} - 1$.
This is, of course, equivalent to the original method,
showing that GHMC effectively satisfies modified detailed balance.

\begin{acknowledgments}
A precursor to this investigation was an exploration of the applicability of
variable-metric HMC~\cite{GiCa11} to particle simulations.
We would like to thank Mari Paz Calvo for her contribution
in bringing that investigation to an end.
The work of JMSS is supported by
Project  MTM2010-18246-C03-01, Ministerio de Ciencia e Innovaci\'on, Spain.
\end{acknowledgments}

%

\appendix

\section{Demonstration of Stationarity}  \label{sss:stat}

Here the variables have one less prime than those of Algorithm~\ref{alg:one}.
The conditional probability density for \substep~(2) is
\begin{equation}  \label{eq:condpd}
\rho(\zee'|\zee) =\theta(\zee)\delta(\zee' -\Psi(\zee))
 + (1 -\theta(\zee))\delta(\zee' - R(\zee))
\end{equation}
where
\[\theta(\zee) =\min\left\{1,
 \frac{\rho(\Psi(\zee))}{\rho(\zee)} |\calJ\Psi(\zee)|
 \right\}\]
if $\zee\in\Omega$ and $\theta(\zee) = 0$ otherwise.
It must be shown that $\int\rho(\zee'|\zee)\rho(\zee)\rmd\zee =\rho(\zee')$.
We have
\begin{eqnarray*}
\int\rho(\zee'|\zee)\rho(\zee)\rmd\zee & = &
 \int_\Omega\min\left\{\rho(\zee),\rho(\Psi(\zee)) |\calJ\Psi(\zee)|\right\}
 \delta(\zee' -\Psi(\zee))\rmd\zee \\
 & &\mbox{} -
 \int_\Omega\min\left\{\rho(\zee),\rho(\Psi(\zee)) |\calJ\Psi(\zee)|\right\}
 \delta(\zee' - R(\zee))\rmd\zee
 + \int\delta(\zee' - R(\zee))\rho(\zee)\rmd\zee.
\end{eqnarray*}
Make the substitution $\zee = R\circ\Psi(\bar{\zee})$ in the first
integral and drop the bars:
\begin{eqnarray*}
\mbox{1st term} & = &
 \int_\Omega\min\left\{\rho\circ R\circ\Psi(\zee),\rho\circ R(\zee)
 |(\calJ\Psi)\circ R\circ\Psi(\zee)|\right\}
 \delta(\zee' - R(\zee))|\calJ(R\circ\Psi)(\zee)|\rmd\zee \\
 & = &
 \int_\Omega\min\left\{\rho\circ R\circ\Psi(\zee)|\calJ(R\circ\Psi)(\zee)|,
 \rho\circ R(\zee))|\calJ(\Psi\circ R\circ\Psi)(\zee)|\right\}
 \delta(\zee' - R(\zee))\rmd\zee.
\end{eqnarray*}
Using Eq.(\ref{eq:grone}),
\[\mbox{1st term} =
 \int_\Omega\min\left\{\rho\circ\Psi(\zee)
|\calJ(R\circ\Psi)(\zee)/(\calJ R)\circ\Psi(\zee)|, \rho(\zee)\right\}
 \delta(\zee' - R(\zee))\rmd\zee,
\]
which cancels the 2nd term.
Also,
\begin{eqnarray*}
\mbox{3rd term} & = &\int
 \delta(\zee' - \zee)\rho\circ R^{-1}(\zee)|\calJ(R^{-1})(\zee)|
 \rmd\zee \\
 & = &\rho(\zee')|(\calJ R)\circ R^{-1}(\zee')||\calJ(R^{-1})(\zee')|
 = \rho(\zee').
\end{eqnarray*}

\section{Derivation of Formulas for Change of Variables}  \label{sss:cv}

Let $A$ be an arbitrary subset of $\bar{\Omega}$. Then
\[\int_A\bar{\rho} =\Pr(\bfw\in A) =\Pr(\zeta^{-1}(\bfz)\in A)
=\int_{\zeta(A)}\rho  =\int_A\rho\circ\zeta\,|\det\partial_w\zeta|,\]
and Eq.~(\ref{eq:cvrho}) follows from
equating integrands and taking their negative logarithms.

To see~(\ref{eq:cvphi}), observe that
\begin{eqnarray*}\Pr(\bfw\in\bar{\Phi}_t(A)) & = &
 \Pr(\zeta^{-1}(\bfz)\in\bar{\Phi}_t(A)) =\Pr(\bfz\in\Phi_t(\zeta(A)))
 =\Pr(\bfz\in\zeta(A)) =\Pr(\zeta^{-1}(\bfz)\in A) \\
 & = &\Pr(\bfw\in A).
\end{eqnarray*}

To verify Eq.~(\ref{eq:cvf}), observe that
 \[\frac{\rmd}{\rmd t}\bar{\Phi}_t
 =\partial_z(\zeta^{-1})\circ\Phi_t\circ\zeta\,
  \frac{\rmd}{\rmd t}\Phi_t\circ\zeta
 = (\partial_w\zeta\circ\zeta^{-1}\circ\Phi_t\circ\zeta)^{-1}\,
   f\circ\Phi_t\circ\zeta
 = (\partial_w\zeta\circ\bar{\Phi}_t)^{-1}\,f\circ\zeta\circ\bar{\Phi}_t.\]

To establish formula~(\ref{eq:cvr}),
 we need to verify Eqs~(\ref{eq:rone}) and~(\ref{eq:rtwo})
 for $\bar{H}$, $\bar{f}$, and $\bar{R}$.
 For Eq.~(\ref{eq:rtwo}),
\begin{eqnarray*}
 \mathit{RHS} & = & -\partial_w(\zeta^{-1}\circ R\circ\zeta)
  \cdot(\partial_w\zeta)^{-1}f\circ\zeta \\
 & = & -\partial_z(\zeta^{-1})\circ R\circ\zeta
  \cdot\partial_z R\circ\zeta\cdot f\circ\zeta \\
 & = & (\partial_w\zeta)^{-1}\circ\zeta^{-1}\circ R\circ\zeta
  \cdot f\circ R\circ\zeta =\mathit{LHS}.
\end{eqnarray*}
 For Eq.~(\ref{eq:rone}), its left-hand side would be
\[\mathit{LHS} =
 H\circ R\circ\zeta -\log|\calJ\zeta\circ\zeta^{-1}\circ R\circ\zeta|
 = H\circ\zeta +\log|\calJ R\circ\zeta|
 -\log|\calJ\zeta\circ\zeta^{-1}\circ R\circ\zeta|. \]
 The right-hand side would be
 \[\mathit{RHS} = H\circ\zeta
 -\log|\calJ\zeta| +\log|\calJ(\zeta^{-1}\circ R\circ\zeta)|.\]
 We have
\[\calJ(\zeta^{-1}\circ R\circ\zeta) =
 (\calJ\zeta\circ\zeta^{-1}\circ R\circ\zeta)^{-1}
 \calJ R\circ\zeta\cdot\calJ\zeta,\]
 whence $\mathit{LHS} =\mathit{RHS}$.

\section{Demonstration of Modified Detailed Balance for a 3-\substep\ Scheme}
  \label{sss:mdetbg}

We have
\[\rho(z'|z) =\int\!\int\rho_3(z'|w')\rho_2(w'|w)\rho_1(w|z)\rmd w\rmd w'\]
where $\rho_3(z'|w') =\rho_1(w'|z')\rho(z')/\rho(w')$.
It is enough to show
\[\frac{\rho(z'|R(z))}{\rho(z')} =\frac{\rho(z|R(z'))}{\rho(z)}.\]
The left-hand side
\begin{eqnarray*}
\mathit{LHS}&=&
\int\!\int\rho_1(w'|z')\rho_1(w|R(z))\frac{\rho_2(w'|w)}{\rho(w')}\rmd w\rmd w'
 \\ &=&\int\!\int
  \rho_1(w'|z')\rho_1(R(w)|R(z))\frac{\rho_2(w'|R(w))}{\rho(w')}|\calJ R(w)|
 \rmd w\rmd w' \\
 &=&\int\!\int
  \rho_1(w'|z')\rho_1(w|z)\frac{\rho_2(w|R(w'))}{\rho(w)}\rmd w\rmd w',
\end{eqnarray*}
where the final step uses Eq.~(\ref{eq:cpdf}).
The right-hand side $\mathit{RHS}$
is this same expression, but with $z$ and $z'$ interchanged.
Equality, $\mathit{RHS}=\mathit{LHS}$, follows by interchanging $w$ and $w'$
in the expression for $\mathit{RHS}$ and applying
modified detailed balance (Eq.~(\ref{eq:gdetbal})).

\section{Demonstration of Modified Detailed Balance for a GHMC Proposal}
  \label{sss:mdetb}

We show that  
the second \substep\ of compressible HMC satisfies modified detailed balance
if the propagator is reversible
and the reversing map $R$ is an involution.
First, consider the case where $z\not\in\Omega$, $z'\not\in\Omega$.
We demonstrate detailed balance, Eq.~(\ref{eq:gdetbal}) separately
for each term of $\rho(\zee'|\zee)$ given by Eq.~(\ref{eq:condpd}).
This means showing
\begin{equation}  \label{eq:dbone}
\theta(R(z))\delta(z' -\Psi(R(z)))\rho(z)
 =\theta(R(z'))\delta(z -\Psi(R(z')))\rho(z')
\end{equation}
and
\[(1 -\theta(R(z)))\delta(z' - z)\rho(z)
 = (1 -\theta(R(z')))\delta(z - z')\rho(z').\]
The second of these obviously holds.
To show the first of these, we write
\begin{eqnarray*}
\delta(z -\Psi(R(z')))
 & = &\delta(\Psi\circ R\circ\Psi\circ R(\zee) -\Psi\circ R(\zee')) \\
& = &\delta(\partial_\zee(\Psi\circ R)(\zee')\cdot(\Psi\circ R(\zee) -\zee'))
 \\
 & = & |\calJ(\Psi\circ R)(\zee')|^{-1}\delta(\zee' -\Psi\circ R(\zee)) \\
 & = & |\calJ(\Psi\circ R)\circ\Psi\circ R(\zee)|^{-1}
\delta(\zee' -\Psi\circ R(\zee)) \\
 & = & |\calJ(\Psi\circ R)|\delta(\zee' -\Psi\circ R(\zee)),
\end{eqnarray*}
where the last step is a consequence of
$1 = \calJ(\Psi\circ R)\circ\Psi\circ R\cdot\calJ(\Psi\circ R)$.
Therefore, to show Eq.~(\ref{eq:dbone}) reduces to showing
\[\theta\circ R\cdot\rho
 = \theta\circ R\circ\Psi\circ R\cdot\rho\circ\Psi\circ R\,|\calJ(\Psi\circ R)|
\]
where omitted arguments are all $z$.
Replacing $z$ by $R(z)$ reduces this to showing
\begin{equation}  \label{eq:dbtwo}
\theta\cdot\rho = \theta\circ R\circ\Psi\cdot\rho\circ\Psi\,|\calJ\Psi|.
\end{equation}
To simplify $\theta\circ R\circ\Psi$, note that
\begin{eqnarray*}
\rho\circ\Psi\circ R\circ\Psi & = &\rho\circ R = \rho/|\calJ R|, \\
\rho\circ R\circ\Psi & = &\rho\circ\Psi/|\calJ R\circ\Psi|, \\
\calJ\Psi\circ R\circ\Psi\cdot\calJ R\circ\Psi\cdot\calJ\Psi & = & \calJ R,
\end{eqnarray*}
whence
\[\theta\circ R\circ\Psi =\min\left\{1,
 \frac{\rho}{\rho\circ\Psi|\calJ\Psi|}
 \right\},\]
from which Eq.~(\ref{eq:dbtwo}) and hence Eq.~(\ref{eq:dbone}) follows.
Now, consider
the case where either $R(z)$ or $R(z')$ is not in $\Omega$.
Without loss of generality, suppose $R(z)\not\in\Omega$.
Suppose $R(z')\in\Omega$.
Then $R\circ\Psi\circ R(z')\in\Omega$ and $R(z)\ne R\circ\Psi\circ R(z')$,
implying $\delta(z -\Psi\circ R(z')) = 0$.
Hence, $\rho(z|R(z')) =\delta(R(z') - R(z))$.
This last equality also holds if $R(z')\not\in\Omega$.
Hence, modified detailed balance reduces to
$\delta(z' - z)\rho(z) =\delta(z - z')\rho(z')$, which always holds.

\end{document}